\documentclass[slac]{revtex4}
\usepackage{graphicx,amsmath,amsfonts}
\usepackage{fancyhdr}
\usepackage{slashed,caption}
\usepackage{braket}

\begin{document}

\title{UNIVERSALITY OF THE WEINBERG THEOREM USING HELICITY CONSTRAINTS}

%

\author{\textbf{Valimbavaka F. Rabearinoro}$^1$}
\author{\textbf{Andriniaina N. Rasoanaivo}$^2$}
\author{\textbf{Roland Raboanary}$^1$}

\affiliation{$^1$Mention Physique et Applications, Domaine des Sciences et Technologies,\\ Antananarivo - BP 906, Universit\'e d'Antananarivo, Madagascar\\
$^2$Sciences Exp\'erimentales et des Math\'ematiques, Ecole Normale Sup\'erieure Ampefiloha,\\
Antananarivo - BP 881,  Universit\'e d'Antananarivo, Madagascar\vspace{.2cm}}
%

\begin{abstract}{
\textbf{\normalsize Abstract:} 
The factorisation of scattering amplitude is described by the Weinberg theorem. In this talk, we will show the universality of the theorem at the next leading correction of the soft expansion. For that we will derive the soft operator by solving helicity constraint relations. }

\end{abstract}

\maketitle

\thispagestyle{fancy}


\section{Introduction}
In quantum physics, the scattering amplitude is the observable connected to the probability for a given process to occur. The amplitudes connect also the theoretical description to the experimental prediction of the large hadron collider. In order to better match these two descriptions we must work on a high precision computation of the amplitude by including low energy corrections.

The low energy behaviour of the amplitude is known to be well described by the Weinberg's theorem \cite{Low:1954kd, Low:1958sn, Weinberg:1965nx, Jackiw:1968zza, Elvang:2016qvq}. In that theorem the low energy limit leads to a factorisation of the scattering amplitude which is represented by the following relation
\begin{equation}
\lim_{s \to 0} A_{n}(1 \cdots , a, s, b, \cdots n) = \hat{S}(a,s^{\pm},b) A_{n-1}(1 \cdots , a, b, \cdots n),
\end{equation}
where $s$ is the label of the soft particle, low energy particle, and $\hat{S}$ is known as the soft operator which is factorised out from the $n$-point amplitude. Such factorisation is known to be universal \cite{Godazgar:2019dkh, Alessio:2019cae, Pate:2019mfs}, i.e. it depends only on the soft particle and the momentums of its adjacents particles, and to simplify its computation it is better to work with its soft expansion
\begin{equation}
\hat{S}=S^{(0)}+S^{(1)}+S^{(2)}+\cdots
\label{expansion}
\end{equation}

This theorem represents the factorisation of the scattering amplitude between the low and the high energy contribution of the amplitude. The factorisation is mainly a separation of the contribution of the particle at low energy $(p_{i}\rightarrow 0)$ and the high energy part which can be useful to include soft radiation correction into the computation.

In this work, our goal is to show the universality of the soft theorem for the case of spin $s=1$ massless particles which are photon and gluon. We will derive the soft operator independently from the amplitude by using helicity constraints as derived in \cite{Rasoanaivo:2020yii}. In section 1, we will derive the first term of the soft operator in the expansion \eqref{expansion} known as the leading soft theorem. In section 2 we will derive the second term in the soft operator expansion known as the next to leading soft. And we will end by a short discussion and conclusion.
\section{LEADING SOFT THEOREM}
\subsection{Helicity constraints}
From the Wigner's little group property of the amplitude, we can derive that the Weinberg factor obey to the following helicity constraints which we aim to resolve
\begin{equation}
[\hat{H}_{i},\hat{S}_{j}]=h_{i}\hat{S}_{j}\delta_{ij}
\label{constraints}
\end{equation}
where $\hat{S}_j$ is the soft operator associated to the $j$-th particle, $h_i$ the helicity of the $i$-th particle, and $\hat{H}_i$ the helicity operator associated to the $i$-th particle such that 
\begin{equation}
\hat{H}_{i}=-\frac{1}{2}\left(\lambda_{i}^{a}\frac{\partial}{\partial \lambda_{i}^{a}} - \bar{\lambda}_{i}^{\dot{a}}\frac{\partial}{\partial \bar{\lambda}_{i}^{\dot{a}}}\right).
\end{equation}

The leading term of the soft operator is also known to be a regular function and the equation \eqref{constraints} can be reduced to 
\begin{equation}
H_{i}S_{j}^{(0)}=
\left \{
\begin{array} {r c l}
&h_{j}S_{j}^{(0)}& i=j \\
&0& i\neq j
\end{array}
\right.
\label{constraints2}
\end{equation}
This above relation is now a linear partial differential equation which can be resolved with boundary conditions.
\subsection{Spinor helicity variables}
Since we are interested in the soft behaviour of massless particle, it is better to work with the spinor helicity formalism which will simplify the resolution of the helicity constraints \eqref{constraints2}. This formalism is based on the isomorphism between Lorentz group of transformation $\mathbb{R}^{1+3}$ and the group of spinors $SL(2,\mathbb C)\otimes SL(2,\mathbb C)$. 
The onshell condition of massless particle, $p^{2}=0$, leads to
\begin{equation}
P_{\mu}\rightarrow P_{a\dot{b}}=\lambda_{a}\bar{\lambda}_{\dot{b}}.
\end{equation}
This means for a momentum $p$ we can associate a square matrix which is equal to the product of a right-handed spinor and a left-handed spinor. With respect to these new variables the new invariant products of two right-handed spinor and two left-handed spinor are represented by the following relations
\begin{equation}
\langle\lambda_{a}\lambda_{b}\rangle=\epsilon^{ab}\lambda_{a}\lambda_{b} \quad\text{ and }\quad
[\bar{\lambda}_{\dot{a}}\bar{\lambda}_{\dot{b}}]=\epsilon^{\dot{a}\dot{b}}\bar{\lambda}_{\dot{a}}\bar{\lambda}_{\dot{b}},
\end{equation}
where $\epsilon$'s are the antisymmetric tensors. 
\subsection{Leading soft}
In this part, we will calculate $S^{(0)}$ which is a function known as the Weinberg's soft factor.  Let be $A_{n}(1, 2, 3, \cdots, n)$ an amplitude and let's consider that the particle $1$ is a soft particle with helicity $h=1$. 
Since the Weinberg's soft factor depends only of the helicity of the soft particle and the momentum of the adjacent particles, therefore we can write
\begin{equation}
S^{(0)}=S^{(0)}(\lambda_{1}, \bar{\lambda}_{1}, \lambda_{2}, \bar{\lambda}_{2}, \lambda_{n}, \bar{\lambda}_{n})
\end{equation}
And we can derive the Weinberg factor by solving the helicity constraints relation $(4)$ 
 \begin{equation}
 \left \{
 \begin{aligned}
 -\frac{1}{2}\left(\lambda_{1}^{1}\frac{\partial}{\partial \lambda_{1}^{1}} + \lambda_{1}^{2}\frac{\partial}{\partial \lambda_{1}^{2}} - \bar{\lambda}_{1}^{\dot{1}}\frac{\partial}{\partial \bar{\lambda}_{1}^{\dot{1}}} - \bar{\lambda}_{1}^{\dot{2}}\frac{\partial}{\partial \bar{\lambda}_{1}^{\dot{2}}}\right)S^{(0)}&=&hS^{(0)}\\
 -\frac{1}{2}\left(\lambda_{2}^{1}\frac{\partial}{\partial \lambda_{2}^{1}} + \lambda_{2}^{2}\frac{\partial}{\partial \lambda_{2}^{2}} - \bar{\lambda}_{2}^{\dot{1}}\frac{\partial}{\partial \bar{\lambda}_{2}^{\dot{1}}} - \bar{\lambda}_{2}^{\dot{2}}\frac{\partial}{\partial \bar{\lambda}_{2}^{\dot{2}}}\right)S^{(0)}&=&0\\
 -\frac{1}{2}\left(\lambda_{n}^{1}\frac{\partial}{\partial \lambda_{n}^{1}} + \lambda_{n}^{2}\frac{\partial}{\partial \lambda_{n}^{2}} - \bar{\lambda}_{n}^{\dot{1}}\frac{\partial}{\partial \bar{\lambda}_{n}^{\dot{1}}} - \bar{\lambda}_{i}^{\dot{2}}\frac{\partial}{\partial \bar{\lambda}_{n}^{\dot{2}}}\right)S^{(0)}&=&0
 \end{aligned}
 \right.
 \end{equation}
  and to simplify the resolution, let's introduce news variables such that:
\begin{equation}
\left \{
\begin{array}{r c l}
x_{1} &=& \langle1 2\rangle,\\
\bar{x}_{1} &=& [1 2],\\
x_{2} &=& \langle1 n\rangle,\\
\bar{x}_{2} &=& [1 n],\\
t &=& \langle2 n\rangle,\\
\bar{t} &=& [2 n].
\end{array}
\right.
\end{equation}
And since the soft factor $S^{(0)}$ should be a Lorentz invariant function and these new variable are made of all possible invariant products, therefore we can write
\begin{equation}
S^{(0)}=S^{(0)}(x_{1}, \bar{x}_{1},x_{2}, \bar{x}_{2}, t, \bar{t} ).
\end{equation}
To resolve the helicity contraints relation for $S^{(0)}$ it is useful to apply variable separation method in which $S^{(0)}=f(x_{1})\bar{f}(\bar{x}_{1})g(x_{2})\bar{g}(\bar{x}_{2})h(t)\bar{h}(\bar{t})$ so the set of partial differential equation becomes
\begin{equation}
\left \{
\begin{array}{r c l}
x_{1}\frac{\partial f(x_{1})}{\partial x_{1}}&=&(a-h+b)f(x_{1})\\
\bar{x}_{1}\frac{\partial \bar{f}(\bar{x}_{1})}{\partial \bar{x}_{1}}&=&(a+h+\bar{b})\bar{f}(\bar{x}_{1})\\
x_{2}\frac{\partial g(x_{2})}{\partial x_{2}}&=&-bg(x_{2})\\
\bar{x}_{2}\frac{\partial \bar{g}(\bar{x}_{2})}{\partial \bar{x}_{2}}&=&-\bar{b}\bar{g}(\bar{x}_{2})\\
t\frac{\partial h(t)}{\partial t}&=&\alpha h(t)\\
\bar{t}\frac{\partial \bar{h}(\bar{t})}{\partial \bar{t}}&=&-\bar{\alpha}\bar{h}(\bar{t})
\end{array}
\right.
\label{equadiff}
\end{equation}
Here $a, b, \bar{b}, \alpha, and \bar{\alpha}$ are the parameters that connect the above individual ordinary differential equations. The resolution of the equations leads to a soft factor that depends on the different parameters,
\begin{equation}
S^{(0)}=\langle1 2\rangle^{a-h+b}\langle1 n\rangle^{-b}\langle2 n\rangle^{\alpha}[1 2]^{a+h+\bar{b}}[1 n]^{-\bar{b}}[2 n]^{-\bar{\alpha}}
\end{equation}
The above function is a solution of the equations \eqref{equadiff} for any value of the different parameters and to be a solution of the helicity constraint \eqref{constraints2}, the parameters are constraint by the following relations
\begin{equation}
\left \{
\begin{array}{r c l}
&-&2h+b+\alpha-\bar{b}+\bar{\alpha}=0\\
&-&b+\alpha+\bar{b}+\bar{\alpha}=0\\
&-&h+b-\bar{b}=0.
\end{array}
\right.
\label{parameter2}
\end{equation}
One of the property of the soft factor is the its pole of order 1 at low energy limit, which can be fixed by dimensional analysis. For a real parameter $\epsilon$, in the soft limit of the momentum $p$ we have:
\begin{equation}
S(\epsilon p^{h_{p}})\overset{\epsilon\rightarrow 0}{\longrightarrow}\frac{1}{\epsilon}S(p^{h_{p}}).
\end{equation}
In order to satisfy such condition the first parameter $a$ is fixed to be equal to $a=-1$.
The resolution of the relation \eqref{parameter2} combined with the soft behaviour of the soft operator, we can derive the expression of $S^{(0)}$ depending on its helicity. 
In one hand, if the helicity of the soft particle is $h=1$, we have $\bar{\alpha}=0$, $\bar{b}=0$, $b=1$ and $\alpha=1$ and the soft factor is 
\begin{equation}
S^{(0)}(1^{+})=\frac{\langle n 2\rangle}{\langle n 1\rangle\langle1 2\rangle}.
\end{equation}
In the other hand the helicity of the soft particle is $h=-1$, we obtain $\alpha=0$, $\bar{\alpha}=-1$, $b=0$ and $\bar{b}=1$ and the soft factor is
\begin{equation}
S^{(0)}(1^{-})=\frac{[n 2]}{[n 1][1 2]}.
\end{equation}

\section{NEXT TO LEADING TERM}
In this part we will derive the first term of correction of the Weinberg factor. It is worth to mention that in the resolution of the soft factor we didn't take into account the momentum conservation of the lower point amplitude when soft contributions are factorized out. Here we will fix this conservation which will generate the next leading contribution of the soft operator.

Let's consider the following soft factorization, where the particle with the momentum $p_1$ is getting soft
\begin{equation}
A_{n}(1, 2, 3 ,\cdots , n)\rightarrow \hat{S}(p_{1})A_{n-1}(2, 3, \cdots, n)
\end{equation}
And if we consider that all the particles are outgoing, the sum of all the momentum of the particles vanishes
\begin{equation}
\sum_{i=1}^{n}p_{i}=0.
\end{equation} 
Therefore the momentum of the lower point amplitude $A_{n-1}$ is not conserved, in which we should have
\begin{equation}
\sum_{i=2}^{n}p_{i}=-p_1.
\end{equation}
In order to make the $n-1$ momentum be conserved, we will shift the adjacent momenta to the soft particle. In fact, for the $n$-point amplitude $A_n$ the momentum conservation is explicitly written as 
\begin{equation}
p_{1}+p_{2}+\cdots+p_{n}=0
\end{equation}
and by shifting $p_2$ and $p_n$ as $p_1$ goes soft so that the new set of $n-1$ momenta will be conserved. Let $\delta_2$ and $\delta_n$ be the shifts such that
\begin{equation}
\left \{
\begin{array}{r c l}
\hat{p}_{2}&=&p_{2}+\delta_{2}\\
\hat{p}_{n}&=&p_{n}+\delta_{n}.
\end{array}
\right.
\end{equation}
In the soft theorem, in order for the momentum to be conserved as the particle of momentum $p_1$ goes soft, a soft factor $S^{(0)}$ will be factorised and the lower point amplitude will depends on the shifted set of momentum

\begin{equation}
\left \{\begin{aligned}
&A_{n}(1, 2, 3 ,\cdots , n)\rightarrow {S}^{(0)}(p_{1})A_{n-1}(\hat{2}, 3, \cdots, \hat{n})\\
&\hat{p}_2+p_3+\cdots+\hat{p}_n=0
\end{aligned}\right .
\end{equation}
This is the same as we apply an operator $\hat{U}(p_1)$ to shift the momenta $p_2$ and $p_n$ in the sub amplitude. Therefore the soft operator can be written as
\begin{equation}
\hat{S}(p_{1})=S^{(0)}(p_{1})\hat{U}(p_{1}).
\end{equation}
In the soft expansion \eqref{expansion} we can relate the next leading of the soft theorem $S^{(1)}$ to be 
\begin{equation}
S^{(1)}(p_{1})=S^{(0)}(p_{1})\hat{T}(p_{1})\quad\text{with}\quad \hat{U}(p_1)=1+\hat{T}(p_1)+\cdots
\end{equation}
Because the process of factorisation of the soft factor $S^{(0)}$ and the momentum shifts are simultaneous, we can impose that them commutation should be 
\begin{equation}
[S^{(0)}(p_{1}),\hat{U}(p_{1})]=[S^{(0)}(p_{1}),\hat{T}(p_{1})]=0.
\label{commutation}
\end{equation}
The main task is now to determine the expression of the shifts $\delta$'s such that the particle 2 and $n$ remain onshell ($\hat{p}_2^2=\hat{p}_n^2$). The onshell condition combined with the momentum conservation leads to two solutions:
\begin{itemize}
\item Case $1$:
 \begin{equation}
 \left \{
 \begin{array}{r c l}
 \hat{p}_{n}&=&\left(\lambda_{n}+\frac{[1 2]}{[n 2]}\lambda_{1}\right)\bar{\lambda}_{n}\\
 \hat{p}_{2}&=&\left(\lambda_{2}+\frac{[n 1]}{[n 2]}\lambda_{1}\right)\bar{\lambda}_{2}
 \end{array}
 \right.
 \end{equation}
 \begin{equation}
 U_{\text{case} 1}(p_{1})=1+\frac{[1 2]}{[n 2]}\lambda_{1}^{a}\frac{\partial}{\partial \lambda_{n}^{a}}+\frac{[1 n]}{[2 n]}\lambda_{1}^{a}\frac{\partial}{\partial \lambda_{2}^{a}}
 \end{equation}
\item Case $2$:
\begin{equation}
\left \{
\begin{array}{r c l}
\hat{p_{n}}&=&\lambda_{n}\left(\bar{\lambda}_{n}+\frac{\langle1 2\rangle}{\langle n 2\rangle}\bar{\lambda}_{1}\right)\\
\hat{p_{2}}&=&\lambda_{2}\left(\bar{\lambda}_{2}+\frac{\langle n 1\rangle}{\langle n 2\rangle}\bar{\lambda}_{1}\right)
\end{array}
\right.
\end{equation}
\begin{equation}
U_{\text{case} 2}(p_{1})=1+\frac{\langle1 2\rangle}{\langle n 2\rangle}\bar{\lambda}_{1}^{\dot{a}}\frac{\partial}{\partial \bar{\lambda}_{n}^{\dot{a}}}+\frac{\langle1 n\rangle}{\langle2 n\rangle}\bar{\lambda}_{1}^{\dot{a}}\frac{\partial}{\partial \bar{\lambda}_{2}^{\dot{a}}}
\end{equation}
\end{itemize}
From the previews section, we found that the expression of the soft factor depends on the helicity of the soft particle, similarly by imposing the commutation \eqref{commutation} we found that the two solutions above $U_{\text{case} 1}$ and $U_{\text{case} 2}$ are respectively the shift operator in the case where the helicity of the soft particle is $h=-1$ and $h=+1$. Which leads to solution 
\begin{equation}
\left \{\begin{aligned}
&S^{(1)}(1^{-})=\frac{[n 2]}{[n 1][1 2]}\left(\frac{\langle1 2\rangle}{\langle n 2\rangle}\bar{\lambda}_{1}^{\dot{a}}\frac{\partial}{\partial \bar{\lambda}_{n}^{\dot{a}}}+\frac{\langle1 n\rangle}{\langle2 n\rangle}\bar{\lambda}_{1}^{\dot{a}}\frac{\partial}{\partial \bar{\lambda}_{2}^{\dot{a}}}\right)\\[.2cm]
&S^{(1)}(1^{+})=\frac{\langle n 2\rangle}{\langle n 1\rangle\langle1 2\rangle}\left(\frac{[1 2]}{[n 2]}\lambda_{1}^{a}\frac{\partial}{\partial \lambda_{n}^{a}}+\frac{[1 n]}{[2 n]}\lambda_{1}^{a}\frac{\partial}{\partial \lambda_{2}^{a}}\right)
\end{aligned}\right .
\end{equation}
\section{Conclusion}
To conclude, we remind that the main goal of this investigation is to show the universality of the Weinberg theorem. And we resolve the helicity constraints to derive the Weinberg factor. 
We derived the soft operator without knowing a particular amplitude. 
The solutions showed us that the soft operators are univeral, ie they depend only the helicity of the soft particle and the momentum of the adjacents particles. 
We have calculated all this in the case of helicity $h=1$ of the soft particle but it can be generalized for any helicity $"h"$.


\begin{thebibliography}{9}
\section*{REFERENCES}
\bibitem{Low:1954kd}
F.~E.~Low,
Phys. Rev. \textbf{96}, 1428-1432 (1954)

\bibitem{Low:1958sn}
F.~E.~Low,
Phys. Rev. \textbf{110}, 974-977 (1958)

\bibitem{Weinberg:1965nx}
S.~Weinberg,
Phys. Rev. \textbf{140}, B516-B524 (1965)

\bibitem{Jackiw:1968zza}
R.~Jackiw,
Phys. Rev. \textbf{168}, 1623-1633 (1968)

\bibitem{Elvang:2016qvq}
H.~Elvang, C.~R.~T.~Jones and S.~G.~Naculich,
Phys. Rev. Lett. \textbf{118}, no.23, 231601 (2017)
[arXiv:1611.07534 [hep-th]].

\bibitem{Godazgar:2019dkh}
H.~Godazgar, M.~Godazgar and C.~N.~Pope,
JHEP \textbf{10}, 123 (2019)
[arXiv:1908.01164 [hep-th]].

\bibitem{Alessio:2019cae}
F.~Alessio and M.~Arzano,
Phys. Rev. D \textbf{100}, no.4, 044028 (2019)
[arXiv:1906.05036 [gr-qc]].

\bibitem{Pate:2019mfs}
M.~Pate, A.~M.~Raclariu and A.~Strominger,
Phys. Rev. D \textbf{100}, no.8, 085017 (2019)
[arXiv:1904.10831 [hep-th]].

\bibitem{Rasoanaivo:2020yii}
A.~N.~Rasoanaivo,
[arXiv:2002.02120 [hep-th]].
\end{thebibliography}
%
%
%
%

\end{document}